# Genocide by Algorithm in Gaza:
# Artificial Intelligence, Countervailing Responsibility, and the Corruption of Public Discourse

Branislav Radeljić

Aula Fellowship for AI, Montreal, Canada
branislav@theaulafellowship.org
ORCID: 0000-0002-0497-3470

**Abstract**
The accelerating militarization of artificial intelligence (AI) has transformed the ethics, politics, and governance of warfare. This article interrogates how AI-driven targeting systems function as epistemic infrastructures that classify, legitimize, and execute violence, using Israel's conduct in Gaza as a paradigmatic case. Building on interdisciplinary debates in security studies, critical technology studies, and international law, the paper advances the concept of "genocide by algorithm" – a mode of technologically mediated extermination in which opaque computational systems automate and rationalize mass violence. The study argues that these systems collapse distinctions between military and civilian targets, and between intentional and accidental harm, thereby eroding the moral and legal frameworks that govern war. Through the lens of responsibility, the article examines three interrelated dimensions: (a) political responsibility, exploring how states exploit AI to accelerate warfare while evading accountability; (b) professional responsibility, addressing the complicity of technologists, engineers, and defense contractors in the weaponization of data; and (c) personal responsibility, probing the moral agency of individuals who participate in or resist algorithmic governance. This is complemented by an examination of the position and influence of those participating in public discourse, whose narratives often obscure or normalize AI-enabled violence. The Gaza case reveals AI not as a neutral instrument but as an active participant in the reproduction of colonial hierarchies and the normalization of atrocity. Ultimately, the paper calls for a reframing of technological agency and accountability in the age of automated warfare. It concludes that confronting algorithmic violence demands a democratization of AI ethics – one that resists technocratic fatalism and centers the lived realities of those most affected by high-tech militarism.

## 1. Introduction

The accelerating integration of artificial intelligence (AI) into military systems[1] raises major concerns across fields such as security studies, international public law, and critical technology studies. As warned in 2023, "[t]he world's major military powers have begun a race to wire AI into warfare. […] No one is inviting AI to formulate a grand strategy, or join a meeting of the Joint Chiefs of Staff. But the same seductive logic that accelerated the nuclear arms race could, over a period of years, propel AI up the chain of command."[2] In the military domain, algorithmic systems function as epistemic infrastructures that order populations into classes of threat, permissible kill, or collateral risk. The opacity of AI systems then becomes a mechanism of governance, which often reconfigures military norms, rendering algorithmic judgments as evidence of objectivity even when they rest on contested data and assumptions.[3] Thus, the "algorithmic gaze" is not simply observational; it is classificatory, prioritizing some lives over others and reproducing existing hierarchies of value (often privileging state violence over civilian protection). Consequently, since AI-driven targeting systems extend the logics of surveillance and "killchain" decision-making, the claim that AI enhances accuracy must be problematized.

Nowhere has this transformation been more evident than in Israel's conduct in Gaza, where AI-driven targeting systems have been used to automate the identification of targets, including individuals, residential blocs, and civil infrastructure. As warned elsewhere, "Israel has constructed GSA [Genocidal Surveillant Assemblage] which utilizes the most advanced technologies of the contemporary period to target Palestinians as a group. It has done so through violating international law for decades and weaponizing illegally collected sensitive information on Palestinians. The GSA is shaped by Israel's settler-colonial project in Palestine and furthers it."[4] For example, the use of AI-assisted systems such as "Gospel," "Laven-

---

[1] L. Gould, M. Hoijtink, M. Jaarsma, & J. Davies, "Innovating algorithmic warfare, experimentation with information manoeuvre beyond the boundaries of the law," *Global Society*, 38(1), (2024), 49–66, https://doi.org/10.1080/13600826.2023.2261466; J. Johnson, "Artificial intelligence & future warfare: implications for international security," *Defense & Security Analysis*, 35(2), (2019), 147–169, https://doi.org/10.1080/14751798.2019.1600800; A. King, "Digital targeting: artificial intelligence, data, and military intelligence," *Journal of Global Security Studies*, 9(2), (2024), ogae009, https://doi.org/10.1093/jogss/ogae009; A.B. Rashid, A.K. Kausik, A.A.H. Sunny, & M.H. Bappy, "Artificial intelligence in the military: an overview of the capabilities, applications, and challenges," *International Journal of Intelligent Systems*, (2023), 1–31, https://doi.org/10.1155/2023/8676366; L. Suchman, "Algorithmic Warfare and the Reinvention of Accuracy," *Critical Studies on Security*, 8(2), (2020), 175–187, https://doi.org/10.1080/21624887.2020.1760587.

[2] R. Andersen, "Never give AI the nuclear codes," *The Atlantic*, 2 May 2023, https://www.theatlantic.com/magazine/archive/2023/06/ai-warfare-nuclear-weapons-strike/673780/. Also, J. Johnson, *AI and the Bomb: Nuclear Strategy and Risk in the Digital Age* (Oxford: Oxford University Press, 2023); D. Matthews, "AI is supposedly the new nuclear weapons: but how similar are they, really?" *Vox*, 29 June 2023, https://www.vox.com/future-perfect/2023/6/29/23762219/ai-artificial-intelligence-new-nuclear-weapons-future; O. Meier, "The fast and the deadly: when artificial intelligence meets the weapons of mass destruction," *European Leadership Network Commentary*, 27 June 2024, https://europeanleadershipnetwork.org/commentary/the-fast-and-the-deadly-when-artificial-intelligence-meets-weapons-of-mass-destruction/.

[3] See, for example, T. Erskine & S.E. Miller, "AI and the decision to go to war: future risks and opportunities," *Australian Journal of International Affairs*, 78(2), (2024), 135–147, https://doi.org/10.1080/10357718.2024.2349598; M. Maas, K. Lucero-Matteucci, & Di Cooke, "Military artificial intelligence as contributor to global catastrophic risk," in *The Era of Global Risk: An Introduction to Existential Risk Studies*, ed. S.J. Beard, M. Rees, C. Richards, & C. Rios-Rojas (Cambridge: Open Book Publishers, 2023), 237–284.

[4] M. Qandeel & Ö. Erdener Topak, "Genocidal surveillant assemblage in Palestine: a socio-legal analysis," *Journal of Genocide Research*, (2025), 1–22, 4, https://doi.org/10.1080/14623528.2025.2567372.
Also, L. Abu-Lughod, "Imagining Palestine's alternatives: settler colonialism and museum politics," *Critical Inquiry*, 47(1), (2020), 1–27, https://doi.org/10.1086/710906; K. Bshara, "Settler colonialism rebranded: Trump's Gaza plan and the capitalist logic of war," *Journal of Palestine Studies*, 54(1), (2025), 62–70, https://doi.org/10.1080/0377919X.2025.2489264;
M. Dumper & A. Badran (Eds.), *Routledge Handbook on Palestine* (Oxon: Routledge, 2025); United Nations, "Report of the Special Rapporteur on the situation of der," "Pegasus," "Iron Dome," and "Where's Daddy," among others, which reportedly automate the identification and prioritization of targets based on opaque and scalable metrics, must be understood within a broader architecture of dehumanization, racial classification, and impunity.[5] This development prompts the concept of "genocide by algorithm," describing a mode of high-tech warfare in which decisions to kill are delegated to opaque computational systems. It also challenges the notion of AI as a neutral instrument; put differently, when embedded in militarized governance structures and unchecked by democratic oversight, AI technologies can facilitate dehumanization and mass atrocities while misrepresenting them as objective and efficient.[6]

This paper addresses three interconnected issues. First, it examines the role of AI as a non-state actor exploited by a state to accelerate warfare and intensify destruction. Second, it digs into the political economy of failed governance that enables crimes to occur without accountability. Third, it seeks to dissect the erosion of critical discourse in a global intellectual climate increasingly shaped by technocratic and securitized logics. While each of these warrants individual analysis, looking at them together, and through the lens of responsibility, offers a deeper understanding of how AI systems collapse distinctions between military and civilian targets, or between intentional and accidental harm, thereby further undermining the legal and moral frameworks that govern warfare. Moreover, the automation of these processes enables state actors to deny intent, shift blame onto machines (as non-state actors), and portray mass casualties as tragic but unintended consequences of algorithmic error. Public discourse, across media, policy, and academia, has often failed to meaningfully engage with (or has actively obscured) the implications of this so-called technological advancement. This contributes to a broader normalization of AI-driven militarism. Taken together, these three issues demonstrate that the case of Gaza exemplifies a profound corruption of both governance and technological ethics.

The paper proceeds with an overview of the main debates surrounding the use of AI in modern warfare, as well as the theoretical and methodological considerations that accommodate the Gaza case, which illustrates a disturbing evolution in military logic and the complicity of both state and non-state actors. Subsequently, the paper considers the three core issues in turn: the role of AI, the responsibility of state and non-state stakeholders, and the position and influence of those participating in public discourse. This analysis calls for a reframing of how we understand technological agency, moral responsibility, and the limitations of legal categories in the age of automated warfare. Together, these issues suggest that the Gaza context reveals not simply a failure of governance or technology, but a systemic condition in which mechanisms of accountability run the risk of being deliberately dismantled. In order words, while discussing a crisis that AI has profoundly intensified, this paper critiques the hidden politics of AI and encourages a rethinking of AI as a participant in public life, as well as the ideological underpinnings of AI systems. The

---

human rights in the Palestinian territories occupied since 1967, Francesca Albanese," 20 October 2025, https://www.un.org/unispal/document/special-rapporteur-report-gaza-genocide-a-collective-crime-20oct25/.
[5] C.H. Gray, *AI, Sacred Violence, and War: The Case of Gaza* (Cham: Palgrave Macmillan, 2025), 79–106. Also, Human Rights Watch, "Questions and answers: Israeli military's use of digital tools in Gaza," 10 September 2024, https://www.hrw.org/news/2024/09/10/questions-and-answers-israeli-militarys-use-digital-tools-gaza; A. Loewenstein, *The Palestine Laboratory: How Israel Exports the Technology of Occupation Around the World* (London: Verso, 2024).
[6] C. Hunter & B.E. Bowen, "We'll never have a model of an AI major-general: artificial intelligence, command decisions, and kitsch visions of war," *Journal of Strategic Studies*, 47(1), (2023), 116–146, https://doi.org/10.1080/01402390.2023.2241648; R. Jenkins, J.P. Sullins, O. Kalu, & K. Phumjam, "Recent insights in responsible AI development and deployment in national defense: a review of literature, 2022–2024," *Journal of Military Ethics*, 24(1), (2025): 63–85, https://doi.org/10.1080/15027570.2025.2483058.

article concludes by highlighting key findings and their potential implications for future policymaking.

## 2. The state of the debate

Conventional security narratives often depict AI as a neutral instrument that enhances precision, reducing "collateral damage."[7] However, this assumption is challenged by the fact that so-called increases in precision or "smart warfare" often mask broader transformations in how violence is authorized and legitimized, including surveillance, remote killing, and the blurring of combatant/civilian distinctions.[8] Accordingly, as some studies insist, "[d]eveloping machine morality is crucial, as we see more autonomous machines integrated into our daily lives,"[9] a task that should not be particularly problematic given findings that "the computational models [artificial conscience and ethical prompting] underlying artificial agents can integrate representations of moral values such as fairness, honesty and avoiding harm."[10] Such efforts are even more pressing in front of the argument pointing out that "while AI can improve efficiency, it may also reduce critical engagement, particularly in routine or lower-stakes tasks in which users simply rely on AI, raising concerns about long-term reliance and diminished independent problem-solving."[11]

From a critical perspective one must interrogate how AI is mobilized in service of unequal power, since "security" is not a neutral category but a political discourse that can justify exceptional violence. There is a growing body of work on how AI and algorithmic systems are changing the nature of warfare, both the conduct of operations and the norms of responsibility; in this field, scholars emphasize that algorithms are sociotechnical systems, not neutral tools; their design, deployment, and effects are shaped by institutional bias, power relations, and political intent.[12] This is amplified by debates about "chatbots' soft power," "political ideology" and "ideological risk governance," as well as "political bias and value misalignment" and "the political legitimacy of global AI governance."[13] The central problem here is: how does one sustain notions of moral responsibility in contexts where decisions are distributed across complex human-machine assemblages?

---

[7] For a debate about benefits and counter-arguments, see Autonomous Weapons, "Myths vs. Fact: Autonomous Weapons and the Global Effort to Control Them," 30 April 2025, https://autonomousweapons.org/autonomous-weapons-misconceptions/; V. Boulanin, L. Saalman, P. Topychkanov, F. Su, & M. Peldán Carlsson, *Artificial Intelligence, Strategic Stability, and Nuclear Risk* (Stockholm: SIPRI, 2020); B.M. Jensen, C. Whyte, & S. Cuomo, "Algorithms at war: the promise, peril, and limits of artificial intelligence," *International Studies Review*, 22(3), (2020), 526–550, https://doi.org/10.1093/isr/viz025; C. Maathuis & K. Cools, "The role of AI in military cyber security: data insights and evaluation methods," *Procedia Computer Science*, 254, (2025), 191–200, https://doi.org/10.1016/j.procs.2025.02.078; J.E. Márquez-Diaz, "Benefits and challenges of military artificial intelligence on the field of defense," *Computación y Sistemas*, 28(2), (2024), 309–323, https://doi.org/10.13053/CyS-28-2-4684.

[8] J. Johnson, "The AI commander problem: ethical, political, and psychological dilemmas of human-machine interactions in AI-enabled warfare," *Journal of Military Ethics*, 21(3-4), (2022), 246–271, https://doi.org/10.1080/15027570.2023.2175887; M. Mayer, "Trusting machine intelligence: artificial intelligence and human-autonomy teaming in military operations," *Defense and Security Analysis*, 39(4), (2023), 521–538. https://doi.org/10.1080/14751798.2023.2264070.

[9] Y. Chu & P. Liu, "Machines and humans in sacrificial moral dilemmas: required similarly but judged differently?" *Cognition*, 239, (2023), 105575, https://doi.org/10.1016/j.cognition.2023.105575.

[10] T. Swoboda & L. Lauwaert, "Can artificial intelligence embody moral values?" *arXiv*, 22 August 2024, https://doi.org/10.48550/arXiv.2408.12250. Also, J. Farnós, A. Sans Pinillos, & V. Costa, "Ethical prompting: toward strategies for rapid and inclusive assistance in dual-use AI systems," *Frontiers in Artificial Intelligence*, 8 (2025), https://doi.org/10.3389/frai.2025.1646444; A. Prettyman, "Artificial conscious-ness," *Inquiry*, (2024), https://doi.org/10.1080/0020174X.2024.2439989.

[11] P. Bois, "Study: relying on artificial intelligence reduces critical thinking skills," *Breitbart*, 12 February 2025, https://www.breitbart.com/tech/2025/02/12/study-relying-on-artificial-intelligence-reduces-critical-thinking-skills/; L. Bai, X. Liu, & J. Su, "ChatGPT: the cognitive effects on learning and memory," *Brain X*, 1(3), (2023), https://doi.org/10.1002/brx2.30; A. Klingbeil, C. Grützner, & P. Schreck, "Trust and reliance on AI: an experimental study on the extent and costs of overreliance on AI," *Computers in Human Behavior*, 160, (2024), 108352. https://doi.org/10.1016/j.chb.2024.108352; C. Zhai, S. Wibowo, & L.D. Li, "The effects of over-reliance on AI dialogue systems on students' cognitive abilities: a systematic review," *Smart Learning Environments*, 11, (2024), https://doi.org/10.1186/s40561-024-00316-7.

This goes hand in hand with the perspective of science and technology studies, which maintains that technologies are not passive conduits of human will, but actors embedded in social order; in other words, they co-produce knowledge, perception, and power. Accordingly, AI systems cannot be treated as neutral tools; they reflect the biases, assumptions, and institutional incentives built into their design, training data, and deployment contexts.[14] Consequently, the involvement of technology complicates legal analyses in terms of the distribution of responsibility in human-machine assemblages, particularly when human oversight is minimal or foregone, when decisions are opaque, or when the architecture allows for plausible deniability.[15] In a system where AI produces lethal outputs, responsibility is fragmented among algorithm designers, data engineers, military controllers, and political leaders; none may fully "own" the action, thereby creating a regime in which shared but diffused responsibility becomes a mechanism of evasion. In effect, AI provides state actors with an opportunity to deflect blame onto algorithmic "errors" while framing civilian deaths as unintended consequences of computational misjudgment. This also means that media and different discourses must grapple with the role of language in framing algorithmic violence in ways that it mutes, naturalizes, or obscures human responsibility (for example, when covering drone strikes, or civilian casualties in counterterror operations).[16]

---

[12] V. Eubanks, *Automating Inequality: How High-Tech Tools Profile, Police, and Punish the Poor* (New York: St. Martin's Press, 2025); M.C. Horowitz, "The ethics & morality of robotic warfare: assessing the debate over autonomous weapons," *Daedalus*, 145(4), (2016), 25–36, https://doi.org/10.1162/DAED_a_00409; A. Nadibaidze, I. Bode, & Q. Zhang, "AI in military decision support systems," Center for War Studies, 2024, https://findresearcher.sdu.dk/ws/portalfiles/portal/275893410/AI_DSS_report_WEB.pdf.

[13] Regarding "chatbots' soft power," see R. Hakimov, D. Rohner, & F. Boboshin, "Chatbots' soft power: generative artificial intelligence promotes different political values across countries," *SSRN*, (2025), 1–28, http://dx.doi.org/10.2139/ssrn.5386746. Regarding "political ideology," see T. Gur, B. Hameiri, & Y. Maaravi, "Political ideology shapes support for the use of AI in policymaking," *Frontiers in Artificial Intelligence*, 30(7), (2024), 1447171, https://doi.org/10.3389/frai.2024.1447171. Regarding "ideological risk governance," see T. Zheng, "Ideological risks in the age of artificial intelligence: realistic representations, generative mechanisms, and response logic," *Journal of Education, Humanities and Social Sciences*, 42, (2024), 685–696, https://doi.org/10.54097/rk2sqq83. Regarding "political bias and value misalignment," see F.Y.S. Motoki, V. Pinho Neto, & V. Rangel, "Assessing political bias and value misalignment in generative artificial intelligence," *Journal of Economic Behavior & Organization*, 234, (2025), 106904, https://doi.org/10.1016/j.jebo.2025.106904. Regarding "the political legitimacy of global AI governance," see E. Erman & M. Furendal, "Artificial intelligence and the political legitimacy of global governance," *Political Studies*, 72(2), (2024), 421–441, https://doi.org/10.1177/00323217221126665.

[14] M.V. Albert, L. Lin, M.J. Spector, & L.S. Dunn (Eds.), *Bridging Human Intelligence and Artificial Intelligence* (Cham: Springer, 2022); V. Capraro, A. Lentsch, D. Acemoglu, et al., "The impact of generative artificial intelligence on socioeconomic inequalities and policymaking," *PNAS Nexus*, 3(6), (2024), 1–18, https://doi.org/10.1093/pnasnexus/pgae191; T. Heyder, N. Passlack, & O. Posegga, "Ethical management of human-AI interaction: theory development review," *Journal of Strategic Information Systems*, 32(3), (2023), 101772, https://doi.org/10.1016/j.jsis.2023.101772; G.M. Johnson, "Algorithmic bias: on the implicit biases of social technology," *Synthese*, 198(10), (2021): 9941–9961. https://doi.org/10.1007/s11229-020-02696-y; T.S. Mullaney, B. Peters, M. Hicks, & K. Philip (Eds.), *Your Computer Is on Fire* (Cambridge, MA: MIT Press, 2021); J. Zeiser, "Owning decisions: AI decision-support and the attributability-gap," *Science and Engineering Ethics*, 30(4), (2024), 1–19, https://doi.org/10.1007/s11948-024-00485-1.

[15] For some aspects of the debate, see N. Anand, "Of code and consequence: assessing the impact of artificial intelligence on international criminal law norms governing the direct and public incitement to genocide," *SSRN* (2024), 1–20, https://dx.doi.org/10.2139/ssrn.5275573; B. Boutin, "State responsibility in relation to military applications of artificial intelligence," *Leiden Journal of International Law*, 36(1), (2023), 133–150, https://doi.org/10.1017/S0922156522000607; O. Müller, "'An eye turned into a weapon:' a philosophical investigation of remote controlled, automated, and autonomous drone warfare," *Philosophy & Technology*, 34(4), (2021), 875–896, https://doi.org/10.1007/s13347-020-00440-5; T.K. Woodcock, "Human/machine(-learning) interactions, human agency and the international humanitarian law proportionality standard," *Global Society*, 38(1), (2024), 100–121, https://doi.org/10.1080/13600826.2023.2267592.

Looking more closely at the genocide in Gaza, accounts focusing on more "traditional" or infrastructural dimensions, including legal aspects, destruction and displacement, and the deprivation of food and water,[17] have been complemented by equally alarming revelations about the use of AI methods. Overall, as some authors have warned, "the war in Gaza represents a structural norm whereby the organized violence is becoming ever more brutal. […] With the ever-increasing expansion of destructive military capacity, the deeper ideological penetration of social order, and the envelopment of micro-level solidarities, contemporary warfare is becoming more detached, dehumanizing, dispassionate, destructive, and ultimately genocidal. In this context, war and genocide can become indistinguishable, and Gaza might be a reliable indicator for what many future violent conflicts could look like."[18] Accordingly, the Israeli military has been accused of committing "AI-assisted genocide,"[19] while tech giants such as Amazon, Alphabet, and Microsoft, have made their AI and cloud infrastructure available to the state of Israel, facilitating activities such as "surveillance," "data processing," and "decision-making," among others.[20] Thus, the concept of "genocide by algorithm" is a conceptual provocation: it pushes us to rethink genocidal violence not solely as a product of intention and actor, but as potentially emergent from engineered systems. The rise of algorithmic violence thus challenges core assumptions of international legal instruments, particularly the requirement of *mens rea* (intent) and clear chain-of-command attribution.

---

[16] K. Chandler, *Unmanning: How Humans, Machines, and Media Perform Drone Warfare* (New Brunswick, NJ: Rutgers University Press, 2020); A.I. Graae & K.Maurer, *Drone Imaginaries: The Power of Remote Vision* (Manchester: Manchester University Press, 2021).

[17] F. Albanese, "Anatomy of a genocide," Human Right Council, 2024, https://www.un.org/unispal/document/anatomy-of-a-genocide/; Amnesty International, "'You feel like you are subhuman:' Israel's genocide against Palestinians in Gaza," 5 December 2024, https://www.amnesty.org/en/documents/mde15/8668/2024/en/; B'Tselem, "Our genocide," (2025), https://www.btselem.org/publications/202507_our_genocide; Human Rights Watch, "Extermination and acts of genocide: Israel deliberately depriving Palestinians in Gaza of water," 19 December 2024, https://www.hrw.org/report/2024/12/19/exterminat0ion-and-acts-genocide/israel-deliberately-depriving-palestinians-gaza; International Association of Genocide Scholars, "IAGS Resolution on the Situation in Gaza," 31 August 2025, https://genocidescholars.org/wp-content/uploads/2025/08/IAGS-Resolution-on-Gaza-FINAL.pdf; O. Bartov, "I'm a genocide scholar, I know it when I see it," *The New York Times*, 15 July 2025, https://www.nytimes.com/2025/07/15/opinion/israel-gaza-holocaust-genocide-palestinians.html; J. Repo, "Genocide and the destruction of the means of social reproduction in Gaza," *European Journal of Politics and Gender*, 8(2), (2025): 492–499, https://doi.org/10.1332/25151088Y2024D000000061.

[18] S. Malešević & L. David, "Organized callousness: Gaza and the sociology of war," *Journal of Genocide Research*, (2025), 1–21, 20, https://doi.org/10.1080/14623528.2025.2503562. Also, Digital Action, "How AI systems are dehumanizing Palestinians," 14 August 2024, https://digitalaction.co/how-ai-systems-are-dehumanising-palestinians/.

[19] Al Jazeera, "'AI-assisted genocide:' Israel reportedly used database for Gaza kill lists," 4 April 2024, https://www.aljazeera.com/news/2024/4/4/ai-assisted-genocide-israel-reportedly-used-database-for-gaza-kill-lists. Also, Y. Abraham, "'Lavender:' the AI machine directing Israel's bombing spree in Gaza," *+972 Magazine*, 3 April 2024, https://www.972mag.com/lavender-ai-israeli-army-gaza/; S. Ishfaq, "Israel's AI-powered genocide of the Palestinians," *Middle East Monitor*, 3 June 2024. https://www.middleeastmonitor.com/20240603-israels-ai-powered-genocide-of-the-palestinians/; L. Katibah, "The genocide will be automated: Israel, AI, and the future of war," Middle East Research and Information Project, 16 October 2024, https://www.merip.org/2024/10/the-genocide-will-be-automated-israel-ai-and-the-future-of-war/.

[20] United Nations, "From economy of occupation to economy of genocide: report of the Special Rapporteur on the situation of human rights in the Palestinian Territories occupied since 1967," 30 June 2025, https://www.un.org/unispal/document/a-hrc-59-23-from-economy-of-occupation-to-economy-of-genocide-report-special-rapporteur-francesca-albanese-palestine-2025/.

More specifically, while the literature on AI's role in genocidal violence is still emergent, important connections are being made, including the notions that (a) AI or algorithmic systems may shift blame or facilitate denial, since state actors sometimes frame civilian deaths as accidents, collateral damage, "errors" in targeting, etc.; (b) algorithmic systems could amplify or automate surveillance, the enforcement of blockades, and decisions about movement/displacement or which civilian infrastructures to spare or destroy; and (c) discourse plays a dual role – to legitimize action and shape international response, largely through the rhetoric of "subhumans" or "enemy civilianization," thus framing civilian casualties as inevitable and contributing to desensitization and normalization. In this context, algorithmic targeting is not merely a matter of technological efficiency; it becomes a tool of systematic erasure. The logics of datafication and automated lethality, when embedded in a settler-colonial structure that already treats Palestinian life as disposable, effectively automate and bureaucratize the destruction of a national group. With this in mind, Gaza should not be treated as an exceptional anomaly but as a particularly intense instantiation of broader tendencies in AI-enabled militarism, which is structured to evade traditional legal accountability while achieving the same destructive ends.

Methodologically, to approach the Gaza question, multiple units of analysis are considered, supported by scholarly literature and primary documents such as policy reports, NGO and expert statements, and investigative journalism and leak-based reporting. Here, it is important to acknowledge that opaque algorithms prevent scrutiny, and many military systems operate under secrecy, which makes it difficult to establish direct causal responsibility from algorithmic decision to civilian harm, or to theorize how much agency one attributes to a system without reifying it as an independent moral actor. While AI systems are never fully autonomous in a vacuum, they may assume roles of decision-making, especially under real-time constraints.[21] This means that confronting "genocide by algorithm" requires more than technical reform or legal innovation – it demands a fundamental reckoning with moral categories, including dignity and accountability, as well as the colonial architectures of technological power and the discursive regimes that sustain them. Here, it is equally important to stress the role of public discourse in helping legitimize or obscure the genocidal implications of AI-enabled warfare, whether through sanitized language, technocratic euphemisms, or the suppression of decolonial and Palestinian voices.

## 3. The perils of (generative) AI

In June 2025, the United Nations Special Rapporteur on the situation of human rights in the Palestinian territories addressed the so-called "dark side of the 'start-up nation,'" confirming, among other things, that "[r]epression of Palestinians has become progressively automated, with tech companies providing dual-use infrastructure to integrate mass data collection and surveillance," and that "[t]he Israeli military has developed AI systems like 'Lavender,' 'Gospel,' and 'Where's Daddy?' to process data and generate lists of targets, reshaping modern warfare and illustrating AI's dual-use nature."[22] Additionally, the report identified IBM, Amazon, Google, Pegasus, Palantir, Microsoft, and Hewlett Packard Enterprises, as providing services that facilitate Israel's actions in Gaza. Knowing the names of any secondary or indirectly involved actors is especially important in light of findings that "four digital tools that the Israeli military is using in Gaza use faulty data and inexact approximations to inform military actions," which prompts further accusations of "Israeli forces violating international humanitarian law, in particular the laws of war concerning distinction between military targets and civilians, and the need to take all feasible precautions before an attack to minimize civilian harm."[23]

---

[21] On this aspect of AI, see M. Albashrawi, "Generative AI for decision-making: a multidisciplinary perspective," *Journal of Innovation & Knowledge*, 10(4), (2025), 100751, https://doi.org/10.1016/j.jik.2025.100751; F.A. Csaszar, H. Ketkar, & H. Kim, "Artificial intelligence and strategic decision-making: evidence from entrepreneurs and investors," *Strategy Science*, 9(4), (2024), 322–345, https://doi.org/10.1287/stsc.2024.0190; B. Sahoh & A. Choksuriwong, "The role of explainable artificial intelligence in high-stakes decision-making systems: a systematic review," *Journal of Ambient Intelligence and Humanized Computing*, 14(6), (2023), 7827–7843, https://doi.org/10.1007/s12652-023-04594-w; L. Wang, Z. Jiang, C. Hu, et al., "Comparing AI and human decision-making mechanisms in daily collaborative experiments," *iScience*, 28(6), (2025), 112711, https://doi.org/10.1016/j.isci.2025.112711.

Understandably, as AI becomes more deeply embedded in weapons systems, the stakes escalate, and AI shifts from a neutral tool to an active participant in geopolitical power plays, amplifying conflict rather than fostering peace or cooperation. With this in mind, and given Israel's settler-colonial approach and deployment of advanced technologies, the case of Palestine represents a sort of "laboratory," where the system in place is best described as the "genocidal surveillant assemblage" or "a new model of genocidal governance by serving as a transcending-amalgamated surveillance infrastructure, gathering all forms of mass surveillance data, fusing and mining these data to assign threat scores, and continuously lowering the threshold for killings and destruction, to achieve the genocidal objective."[24] For others, Israel's use of AI for military purposes, which represents a "postmodern asymmetric war," is far from straightforward, mainly because its success depends on "completing the ethnic cleansing of Palestinians."[25] Consequently, apart from speaking of "practices [that] are certainly doomed to fail," as they stem from "victory-through-technology illusions" and "immoral and ineffectual military operations," it is also impossible to exclude "mass casualty events in Israel," as per the war's postmodern asymmetric nature.[26]

In this context, the United Nations' claims that "Israel-located servers ensure data sovereignty and a shield from accountability," and that its Pegasus spyware "used against Palestinian activists and licensed globally to target leaders, journalists and human rights defenders," and, perhaps most importantly, "enables 'spyware diplomacy' while reinforcing state impunity,"[27] are closely related to the question of AI and state sovereignty. In other words, AI can be viewed as both a state and a non-state actor, or as a sophisticated mixture of the two, which complicates the debate about accountability. While closely related to, if not a direct offspring of, major tech giants, its regulation and deployment remain largely state-dependent matters. This helps explain why the Washington administration – and the US is admittedly at the forefront of AI developments – maintains strong ties with big tech, elevating their activities ever closer to the state level and moving them away from a purely non-state status. Thus, when scholars argue that non-state actors can contribute to "the fragmentation of political responsibility" and that "[t]he more successful non-state actors are in affecting political outcomes, the more responsibility they should be asked to take for those outcomes,"[28] it becomes clear that AI is being instrumentalized. This further undermines claims about AI's neutrality, as its supposed neutrality may in fact translate into silence and heightened control, where inclusion is not about genuine representation or the participation of all humanity, but rather about algorithm-driven submission and the reinforcement of existing hierarchies.

---

[22] United Nations, "From economy of occupation to economy of genocide." Also, Gray, *AI, Sacred Violence, and War*, 79–106; M.O. Jones, "In Gaza, we are watching the world's first AI-assisted genocide," *The New Arab*, 10 April 2024, https://www.newarab.com/analysis/israel-carrying-out-ai-assisted-genocide-gaza; M. Kwet, "How US big tech supports Israel's AI-powered genocide and apartheid," *Al Jazeera*, 12 May 2024, https://www.aljazeera.com/opinions/2024/5/12/how-us-big-tech-supports-israels-ai-powered-genocide-and-apartheid; R. Mohydin, "Israel's AI-powered genocide raises new moral perils," TRT World Research Centre, 23 April 2024, https://researchcentre.trtworld.com/publications/policy-outlook/israels-ai-powered-genocide-raises-new-moral-perils/; A. Rahman, "Explainer: the role of AI in Israel's genocidal campaign against Palestinians," Institute for Palestine Studies, 16 October 2024, https://www.palestine-studies.org/en/node/1656285.
[23] Human Rights Watch, "Questions and answers;" Loewenstein, *The Palestine Laboratory*. Also, B. McKernan & H. Davies, "'The machine did it coldly:' Israel used AI to identify 37,000 Hamas targets," *The Guardian*, 3 April 2024, https://www.theguardian.com/world/2024/apr/03/israel-gaza-ai-database-hamas-airstrikes; Sam Mednick, Garance Burke, and Michael Biesecker, "How US Tech Giants Supplied Israel with AI Models, Raising Questions about Tech's Role in Warfare," *Associated Press*, 18 February, 2025, https://apnews.com/article/israel-palestinians-ai-weapons-430f6f15aab420806163558732726ad9; N. Sylvia, "The Israel defense forces' use of AI in Gaza: a case of misplaced purpose," *RUSI Commentary*, 4 July 2024, https://www.rusi.org/explore-our-research/publications/commentary/israel-defense-forces-use-ai-gaza-case-misplaced-purpose; United Nations, "Gaza: UN experts deplore use of purported AI to commit 'domicide' in Gaza, call for reparative approach to rebuilding," Office of the High Commissioner, 15 April 2024, https://www.ohchr.org/en/press-releases/2024/04/gaza-un-experts-deplore-use-purported-ai-commit-domicide-gaza-call.
[24] Qandeel & Erdener Topak, "Genocidal Surveillant Assemblage," 21.
[25] Gray, *AI, Sacred Violence, and War*, 8.
[26] Ibid., 6–7.
[27] United Nations, "From economy of occupation to economy of genocide."

As underscored by a study looking at China, the European Union, and the United States, "[AI] standards are neither objective nor neutral" since "[they] are crafted by actors who make normative choices in particular institutional contexts, subject to political and economic incentives and constraints," which invalidates the ambition of having "a single, best formula to disseminate across contexts."[29] In other words, the so-called AI race risks prioritizing global dominance over responsible, inclusive innovation, whereby the illusion of AI neutrality becomes politically dangerous, reinforcing authoritarianism and undermining democratic deliberation.[30] As recent studies have warned, AI platforms (and billionaires in charge) possess the power to influence national security and public discourse in ways that bypass public oversight, potentially amplifying exclusion and surveillance, and prioritizing strategic gains over collective well-being.[31] This dynamic speaks to AI's elevated status, reflecting that of a state rather than a non-state actor. It also highlights how the

---

[28] A. Sabl, "Governing pluralism," in *Public Ethics and Governance: Standards and Practices in Comparative Perspective*, ed. D. Saint-Martin & F. Thompson (Stamford, CT: JAI Press, 243–256), 250.

[29] A. Solow-Niederman, "Can AI standards have politics?" *UCLA Law Review*, 71, (2023), 230–245, 230, https://www.uclalawreview.org/can-ai-standards-have-politics/. Also, S. Chowdhury & L. Klautzer, "Shaping an adaptive approach to address the ambiguity of fairness in AI: theory, framework, and illustrations," *Cambridge Forum*, 1, (2025), 1–17, https://doi.org/10.1017/cfl.2025.7; S. Fazelpour & D. Danks, "Algorithmic bias: senses, sources, solutions," *Philosophy Compass*, 16(8), (2021): e12760, https://doi.org/10.1111/phc3.12760; H. Wang & V. Blok, "Why putting artificial intelligence ethics into practice is not enough: towards a multi-level framework," *Big Data & Society*, 12(2), (2025), 1–14, https://doi.org/10.1177/20539517251340620; G.I. Zekos, *Political, Economic and Legal Effects of Artificial Intelligence: Governance, Digital Economy and Society* (Cham: Springer, 2022).

[30] A. Arora, M.I. Barrett, E. Lee, et al., "Risk and the future of AI: algorithmic bias, data colonialism, and marginalization," *Information and Organization*, 33(3), (2023), 100478, https://doi.org/10.1016/j.infoandorg.2023.100478; D. Bernstein, "Who is winning the AI arms race?" 28 August 2024, https://www.forbes.com/sites/drewbernstein/2024/08/28/who-is-winning-the-ai-arms-race/; N.A. Smuha, "From a 'race to AI' to a 'race to AI regulation:' regulatory competition for artificial intelligence," *Law, Innovation, and Technology*, 13(1), (2021), 57–84, https://doi.org/10.1080/17579961.2021.1898300; J. Sublime, "The AI race: why current neural network-based architectures are a poor basis for artificial general intelligence," *Journal of Artificial Intelligence Research*, 79, (2023), 41–67, https://doi.org/10.1613/jair.1.15315; J. Thornhill, "The AI race is generating a dual reality," *The Financial Times*, 18 April 2024, https://www.ft.com/content/8af1f467-2953-4cbc-a336-4c92c92e6792.

[31] B. Gardiner, "How Silicon Valley is disrupting democracy," *MIT Technology Review*, 13 December 2024, https://www.technologyreview.com/2024/12/13/1108459/book-review-silicon-valley-democracy-techlash-rob-lalka-venture-alchemists-marietje-schaake-tech-coup/amp/; T. McKenzie, "Twitter/X will use your posts for AI training and there's no opting out," *80 Level*, 17 October 2024, https://80.lv/articles/twitter-x-will-use-your-posts-for-ai-training-and-there-s-no-opting-out/.

Washington administration's decision to supplement its traditional support for the Israeli state (in terms of politics, financial assistance, and supply of military)[32] with US-originating AI systems raises a crucial question of complicity, not only of AI itself but also of the United States in Israel's genocide in Gaza. This, in turn, inevitably opens another question – that of regulatory mechanisms,[33] which are often tied to the overall context, including to public–private (state–non-state) arrangements and to those responsible for high-stakes decisions operating under commercial or governmental pressure. When it comes to the United States, as succinctly explained elsewhere, "[t]he spread of authoritarianism in the political realm is paralleled by trends in the American technology sector, where CEOs increasingly exercise authority without meaningful structural oversight or accountability."[34] With this in mind, bringing the status of AI closer to that of the state suggests a partnership driven by a particular ideological orientation. Put differently, and in relation to the debate on US complicity in "genocide by algorithm" in Gaza, the argument that "[s]tate responsibility can be engaged for the wrongful use of AI-enabled technologies in the battlefield, negligent development or procurement of AI technologies […], and failure to ensure respect for international law by other actors developing or deploying AI,"[35] is of utmost relevance and warrants further examination.

## 4. The puzzling nature of countervailing responsibility

Policymakers – either as content creators, involved in formulation, implementation, and amendments to policies, or as individuals with the authority to sign off decisions, approve policy documents, and determine allocation of financial resources – represent key stakeholders in the process. However, given that they are also deeply invested in preserving their positions of power and securing future electoral or political advantages, their handling of policy proposals, including the decision to approve, delay, or reject them and their subsequent formalization through institutional procedures, cannot be separated from a complex matrix of political, strategic, and often personal calculations.[36] In parallel, the concept of countervailing responsibility has become increasingly relevant in scholarly debates, especially as governance grows more complex and multi-layered. Back in the 1980s, Michael Harmon identified three distinct types of administrative responsibility – political, professional, and personal – and highlighted the inherent tensions among them. As noted, "[a]ction that is deemed correct from the standpoint of one meaning might very well be incorrect or irresponsible from the standpoint of another."[37] More precisely, while political responsibility implies "[a]ction that is accountable to or consistent with objectives or standards of conduct mandated by political or hierarchical authority," professional responsibility implies "[a]ction that is informed by professional expertise, standards of ethical conduct, and by experience rooted in agency history and traditions," and personal responsibility implies "[a]ction that is informed by self-reflexive understanding; and emerges from a context of authentic relationships wherein personal commitments are regarded as valid bases for moral action."[38] These overlapping, and at times competing, forms of responsibility reflect the ethical and operational dilemmas that public officials frequently navigate.

---

[32] See, for example, A.I. Roth, "Reassurance: a strategic basis of US support for Israel," *International Studies Perspectives*, 10(4), (2009), 378–393, https://doi.org/10.1111/j.1528-3585.2009.00384.x; J.J. Mearsheimer & S.M. Walt, *The Israel Lobby and US Foreign Policy* (London: Penguin, 2007); E. Shakman Hurd, "American support for Israel is a political religion," *New Lines Magazine*, 15 September 2025, https://newlinesmag.com/argument/american-support-for-israel-is-a-political-religion/.

[33] For a good debate regarding regulation, see R. Medaglia, J.R. Gil-Garcia, & T.A. Pardo, "Artificial intelligence in government: taking stock and moving forward," *Social Science Computer Review*, 41(1), (2021), 123–140, https://doi.org/10.1177/08944393211034087; G. Papyshev & M. Yarime, "The state's role in governing artificial intelligence: development, control, and promotion through national strategies," *Policy Design and Practice*, 6(1), (2023), 79–102. https://doi.org/10.1080/25741292.2022.2162252; A. Taeihagh, "Governance of artificial intelligence," *Policy & Society*, 40(2), (2021), 137–157. https://doi.org/10.1080/14494035.2021.1928377; S. Tomić & V. Štimac, "Pinning down an octopus: towards an operational definition of AI systems in the EU AI Act," *Journal of European Public Policy*, (2025), 1–36, https://doi.org/10.1080/13501763.2025.2534648.

[34] G.F. Davis, "Authoritarianism with Silicon Valley characteristics," *Journal of Management Inquiry*, 32(1), (2022), 3–20, 7, https://doi.org/10.1177/10564926221119395. Also, M. Dean, "The concept of authoritarian governmentality today," *Global Society*, 39(1), (2025), 16–35, https://doi.org/10.1080/13600826.2024.2362739; A.E. dos Reis Peron, D. Almstadter Mattar de Magalhães, & G. Fernandes Caetano, "Beyond digital repression: techno-authoritarian radical rights governments," *Cogent Social Sciences*, 11(1), (2025), 2528457, https://doi.org/10.1080/23311886.2025.2528457.

[35] Boutin, "State responsibility," 149.

[36] For a good overview, see P.J. May, "Politics and policy analysis," *Political Science Quarterly*, 101(1), (1986), 109–125, https://doi.org/10.2307/2151446.

Regarding political responsibility, which Harmon sees as "hierarchical" and "synonymous with accountability," and which is "enforceable through legal sanctions" and complemented with "the notion of efficiency, […] defined as the ratio of expenditures to results,"[39] the claim that Israel has used algorithmic or AI systems to conduct operations in Gaza that amount to genocide raises three major considerations. First, at the level of the Israeli government and military hierarchy, the political objective is to defeat Hamas, but if the means involve large-scale killing of civilians through algorithmic targeting, then the political responsibility dimension becomes crucial. Israel is accountable (or should be held accountable) to international norms (the Genocide Convention or international humanitarian law) for the conduct of its armed forces. Accordingly, the use of AI to enable genocidal acts, can be interpreted in two, equally disturbing ways – that the algorithmic deployment is not aligned with norms of conduct (and thus it is unlawful), and since AI is a state-supported project, which resulted in mass destruction of civilian life, that the state has failed its political responsibility.[40] Second, at the level of big tech companies embedded in the US political economy, though they are private actors, they operate under state regulation, benefit from state support, and often coordinate with governmental intelligence/military systems. Even if the political dimension here is ignored, if companies are effectively arms of the state, then their actions fall into the realm of political responsibility; put differently, if these companies assist or enable genocide through algorithmic systems, they share in political responsibility by virtue of their alignment with state-mandated action.[41] Third, at the level of the US and any other foreign government, their political responsibility lies in what they enable or support, including military aid, intelligence sharing, and technology transfers. In other words, if the hierarchical apparatus of foreign administrations permits Israeli actions or algorithmic infrastructure that facilitate genocide, then they have political responsibility.[42]

---

[37] M.M. Harmon, "The responsible actor as 'tortured soul:' the case of Horatio Hornblower," *Administration & Society*, 21(3), (1989), 283–312, 286, https://doi.org/10.1177/009539978902100301.
[38] Ibid., 289.
[39] Ibid., 290.
[40] See, for example, A. Abou Shhadeh, "Israel's moral compass is smashed: the state's only language is force," *Middle East Eye*, 30 August 2025, https://www.middleeasteye.net/opinion/gaza-genocide-israels-moral-compass-smashed-states-only-language-force; Y. Abraham, "Leaked documents expose deep ties between Israeli army and Microsoft," *+972 Magazine*, 23 January 2025, https://www.972mag.com/microsoft-azure-openai-israeli-army-cloud/; E. Banco, J. Landay, & H. Pamuk, "Exclusive: US intel found Israeli military lawyers warned there was evidence of Gaza war crimes," *Reuters*, 7 November 2025, https://www.reuters.com/world/us/us-intel-found-israeli-military-lawyers-warned-there-was-evidence-gaza-war-2025-11-07/; M. Biesecker, S. Mednick, & G. Burke, "As Israel uses US-made AI models in war, concerns arise about tech's role in who lives and who dies," *The Associated Press*, 18 February 2025, https://apnews.com/article/israel-palestinians-ai-technology-737bc17af7b03e98c29cec4e15d0f108.
[41] See, for example, O. Ahmed, "Israel's influence on big tech: silencing pro-Palestine media," *Middle East Monitor*, 23 August 2024, https://www.middleeastmonitor.com/20240823-israels-influence-on-big-tech-silencing-pro-palestine-media/?amp; L. Fang & J. Poulson, "Pro-Israel group censoring social media led by former Israeli intelligence officers," 12 July 2024, https://www.leefang.com/p/pro-israel-group-censoring-social; B. Perrigo, "Exclusive: Google contract shows deal with Israel Defense Ministry," *Time*, 12 April 2024, https://time.com/6966102/google-contract-israel-defense-ministry-gaza-war/.

Regarding professional responsibility, which Harmon sees as "the avoidance of wrongdoing, especially action intended to reap personal or political gain," and which goes hand in hand with the notions of "opportunism" and "goal displacement,"[43] in the context of Israel's "genocide by algorithm" in Gaza, professional responsibility applies to military planners, tech corporations, algorithm designers, intelligence analysts, and regulatory agencies. While it is reasonable to believe that in Israel's military and intelligence apparatus, there are professionals who must act in accordance with professional standards (training, international law, and ethical codes of military conduct), it is equally important to state that when such professionals either design or approve systems which enable automated targeting and mass civilian casualties, they breach the professional responsibility.[44] Similarly, big tech companies, including their professionals in AI governance, machine learning, and systems engineering, carry professional responsibility to ensure their algorithms are not used to facilitate human rights violations.[45] Failure to do so means they are complicit via professional neglect. Finally, US professionals, including regulators, military lawyers, and intelligence ethicists, also hold professional responsibility to ensure that algorithmic systems provided to allies (for example, Israel) or developed by US-based companies do not facilitate mass extermination or systematic destruction of populations.[46] If they fail to enforce or monitor such systems, they are breaching professional standards.

Regarding personal responsibility, which Harmon sees as "an awareness of both the existence and the social and psychological limitations of one's role as agent or cause of one's action,"[47] and which should also be considered vis-à-vis the notions of "impersonality," "narcissism," and "confluence," the involvement of AI in atrocities in Gaza accounts for the moral agency of algorithms and, perhaps even more, that of individuals such as political leaders, military commanders, data scientists, software engineers, and corporate executives – each one expected to decide whether to act or resist. For example, a military commander in Israel who approves an AI-targeting algorithm has personal responsibility to step back and reflect if that is consistent with my moral commitments, my understanding of human rights and humanity.[48] If they simply follow orders or if they actively enable algorithmic destruction without reflexive moral consideration, they breach personal responsibility. Similarly, a big tech engineer working at a company whose algorithm is used for targeting civilians in Gaza has a personal responsibility to act with moral commitment, resist or blow the whistle if the system facilitates genocide. If they remain silent or complicit, they fail their personal responsibility.[49] Finally, a US congressperson or any other government authorizing military aid, or a civilian oversight official, also holds personal responsibility as to whether their actions align with moral commitments to protect human life.[50] Their silence or support may constitute personal complicity.

---

[42] See, for example, B. Elmalı, "Quo vadis, Western order," *Anadolu Agency*, 2 January 2024, https://www.aa.com.tr/en/analysis/opinion-quo-vadis-western-order/3097807; A. Loewenstein, "Israel's use of AI in Gaza is a terrifying model coming to a country near you," *Middle East Eye*, 28 January 2025, https://www.middleeasteye.net/opinion/israel-use-ai-gaza-terrifying-model-coming-country-near-you; M. Maggiore, "Lawyers file case against EU Commission and Council for 'failure to act' on Gaza genocide," *EUobserver*, 15 July 2025, https://euobserver.com/eu-and-the-world/ar48acfd1a.

[43] Harmon, "The responsible actor," 290–294.

[44] See, for example, A. Downey, "The alibi of AI: algorithmic models of automated killing," *Digital War*, 6(6), (2025), 1–19, https://doi.org/10.1057/s42984-025-00105-7; S. Düz & M.S. Koçakoğlu, "Deadly algorithms: destructive role of artificial intelligence in Gaza War," SETA, 2025, https://media.setav.org/en/file/2025/02/deadly-algorithms-destructive-role-of-artificial-intelligence-in-gaza-war.pdf; W. Wells, "Battlefield evidence in the age of artificial intelligence-enabled warfare," *Chicago Journal of International Law*, 26(1), (2025), 249–280, https://cjil.uchicago.edu/print-archive/battlefield-evidence-age-artificial-intelligence-enabled-warfare.

[45] See, for example, Arab Center for the Advancement of Social Media, "Palestinian digital rights, genocide, and big tech accountability," 2024, https://7amleh.org/storage/genocide/English%20new%20(1).pdf; Z. Motala, "Big tech's complicity in genocide: the unforgivable silence of online platforms," *Middle East Monitor*, 21 September 2024, https://www.middleeastmonitor.com/20240921-big-techs-complicity-in-genocide-the-unforgivable-silence-of-online-platforms/; E. Sype, "Big tech terror: for Palestinians, AI apocalypse is already here," *The New Arab*, 24 July 2024, https://www.newarab.com/opinion/big-tech-terror-palestinians-ai-apocalypse-already-here.

[46] See, for example, L.J. Bilmes, W.D. Hartung, & S. Semler, "United States spending on Israel's military operations and related US operations in the region, October 7, 2023 – September 30, 2024," Watson Institute for International & Public Affairs, 2024, https://watson.brown.edu/costsofwar/files/cow/imce/papers/2023/2024/Costs%20of%20War_US%20Support%20Since%20Oct%207%20FINAL%20v2.pdf; W.D. Hartung, "US military aid and arms transfers to Israel, October 2023–September 2025," Quincy Institute for Responsible Statecraft, 7 October 2025, https://quincyinst.org/research/u-s-military-aid-and-arms-transfers-to-israel-october-2023-september-2025/#.

[47] Harmon, "The responsible actor," 300–306.

      Together, the above responsibilities are essential because the scale of technological systems can diffuse accountability; the algorithm acts as a state, and it may be abused to diminish individual agency or agency of the state. In any case, the question of "genocide by algorithm" prompts the following question of who among the individual actors had the moral capacity to resist, to question the algorithmic deployment, to refuse to participate? This deepens the argument by highlighting that the system is not simply structural but staffed by individuals who make choices (or fail to make them) and thus bear personal moral responsibility for complicity in genocide. The tension is acute; an individual may have professional/technical duty (professional responsibility) or political/hierarchical duty (political responsibility) but may see that duty conflicting with their personal moral commitments. Harmon's insight is precisely that these responsibilities can conflict – what is correct politically may be wrong personally, or what is correct professionally may violate personal morality.[51] Accordingly, the argument presented here is not just a structural critique ("the state did this") but a layered account of responsibility across hierarchical, institutional, and personal levels, revealing how "genocide by algorithm" emerged from interlocking failures.

## 5. The public between protest and self-censorship

Large language models (LLMs) are capable of producing "valid position estimates, [which] can be used, in turn, to position political actors such as politicians;" for example, "with individual tweets, ancillary analyses show that the position estimates returned by the best LLMs are as accurate as the average of the ratings of [multiple] independent human coders."[52] Thus, even if (generative) AI would like to come across as neutral or apolitical when asked uncomfortable questions, the power of platforms such as Twitter/X, which is used to train AI,[53] is limitless when deciding whom to cancel and whom to grant the right to freedom of expression. This corresponds to the point regarding standards and global governance: "[T]ailoring a standard to local conditions and accounting for the political economy and normative commitments of a particular context may make it impossible to craft a standard that would be well suited for broader diffusion. […] At the bottom, there is a real question of whether it is possible to embrace the politics of standards and still have them meaningfully serve as standards."[54] In other words, AI is far more than a technical innovation – it is a sociotechnical force that both reflects and reproduces ideological, geopolitical, and economic power structures. Thus, as a discursive technology, it implies a system that not only performs functions but also channels voice, organizes meaning, and regulates participation in the public sphere. Consequently, AI must be understood not only as a site of technological advancement but as a terrain of political struggle.

---

[48] See, for example, S. Düz, "Gaza as a testing ground: Israel's AI warfare," *SETA*, 3 July 2025, https://www.setav.org/en/gaza-as-a-testing-ground-israels-ai-warfare; B. Elmalı, "Israel has tainted AI with genocide," *Anadolu Agency*, 1 May 2024, https://www.aa.com.tr/en/analysis/opinion-israel-has-tainted-ai-with-genocide/3207242.

[49] See, for example, M. Biesecker, "Microsoft workers protest sale of AI and cloud services to Israeli military," *The Associated Press*, 25 February 2025, https://apnews.com/article/israel-palestinians-ai-technology-microsoft-gaza-lebanon-90541d4130d4900c719d34eb cd67179d; J. Novet, "Microsoft letting employees raise concerns about products after Middle East controversy," *CNBC*, 5 November 2025, https://www.cnbc.com/2025/11/05/microsoft-offers-employees-trusted-technology-review-form.html; M. Sainato, "Workers accuse Google of 'tantrum' after 50 fired over Israel contract protest," *The Guardian*, 27 April 2024, https://www.theguardian.com/technology/2024/apr/27/google-project-nimbus-israel.

[50] See, for example, L. Mascaro, "House approves nearly $14.5 billion in military aid for Israel: Biden vows to veto the GOP approach," *The Associated Press*, 3 November 2023, https://apnews.com/article/house-israel-aid-ukraine-repu blicans-biden-gaza-b7bfe528b12ac5954cfd5c034f11320d; T. Perkins, "Revealed: Congress backers of Gaza War received most from pro-Israel donors," *The Guardian*, 10 January 2024, https://www.theguardian.com/us-news/2024/jan/10/congress-member-pro-israel-donations-military-support.

[51] For an in-depth debate, see M.M. Harmon, *Responsibility as Paradox: A Critique of Rational Discourse on Government* (Thousand Oaks, CA: SAGE, 1995).

The increasing alignment between political and corporate elites has blurred the boundaries between public interest and private gain. As power continues to concentrate, the influence of unelected actors has become impossible to ignore. As noted elsewhere, "[b]illionaires are now effectively making national security decisions on behalf of the United States and using their social media companies to push right-wing agitprop and conspiracy theories."[55] While AI may be programmed to be aware of the concerns of all, its functionality is inseparable from the capital invested in its development and the political or corporate interests influencing its priorities. The ideological underpinnings of its architecture cannot be separated from the intentions and incentives of those who fund, train, and deploy it.[56] Democratic systems, once thought to provide the tools to hold elected leaders accountable, now appear increasingly fragile under the weight of surveillance capitalism, data monopolies, and digital authoritarianism.[57] In such a context, where strikes, protests, and boycotts are neutralized by legal or algorithmic means, the very notion of the "informed voter" begins to unravel. Instead, citizens are reimagined not as agents of change, but as passive facilitators of a system they cannot meaningfully influence. Eventually, the outcome is chilling: once transformed into subjects of soft control – nudged by algorithms, distracted by curated feeds, and overwhelmed by noise – voters may end up

---

[52] G. Le Mens & A. Gallego, "Positioning political texts with large language models by asking and averaging," *Political Analysis*, 33(3), (2025), 274–282, 280, https://doi.org/10.1017/pan.2024.29. Also, Collective Intelligence Project, "White Paper," (2024), https://www.cip.org/whitepaper; A. Jungherr, "Artificial intelligence and democracy: A conceptual framework," *Social Media + Society*, 9(3), (2023), 1–14, https://doi.org/10.1177/20563051231186353.
[53] McKenzie, "Twitter/X will use your posts."
[54] Solow-Niederman, "Can AI standards," 244. Also, Zekos, *Political, Economic and Legal Effects*; J. Tallberg, E. Erman, M. Furendal, et al., "The global governance of artificial intelligence: next steps for empirical and normative research," *International Studies Review*, 25(3), (2023), viad040, https://doi.org/10.1093/isr/viad040.
[55] Gardiner, "How Silicon Valley."
[56] C. Coglianese & D. Lehr, "Transparency and algorithmic governance," *Administrative Law Review*, 71(1), (2019), 1–56, https://scholarship.law.upenn.edu/faculty_scholarship/2123; M. Ebers & M. Cantero Gamito (Eds.), *Algorithmic Governance and Governance of Algorithms: Legal and Ethical Challenges* (Cham: Springer, 2021); I. Kalpokas, *Algorithmic Governance: Politics and Law in the Post-Human Era* (Cham: Palgrave Macmillan, 2019); R. Khan, T. Mackenzie, S. Bhaduri, et al., "Whole-person education for AI engineers," *arXiv*, 10 June 2025, https://doi.org/10.48550/arXiv.2506.09185; S. Srivastava, "Algorithmic governance and the International Politics of Big Tech," *Perspectives on Politics*, 21(3), (2023), 989–1000, https://doi.org/10.1017/S1537592721003145.
[57] T. Roberts & M. Oosterom, "Digital authoritarianism: a systematic literature review," *Information Technology for Development*, 31(4), (2025), 860–884, https://doi.org/10.1080/02681102.2024.2425352; S. Zuboff, *The Age of Surveillance Capitalism: The Fight for a Human Future at the New Frontier of Power* (London: Profile Books, 2019).

enabling the very systems that undermine their autonomy.

Looking at the governance realm more broadly, (generative) AI is not only a productive technology of knowledge but also a modality of power, which echoes Michel Foucault's argument that modern power operates not through overt coercion, but through subtle processes of normalization, surveillance, and self-regulation – now accelerated by AI infrastructures.[58] AI technologies are increasingly integrated into institutions where they structure what can be seen, known, and acted upon; whether in predictive policing, algorithmic risk assessments in the criminal justice system, or automated eligibility determinations in healthcare, AI functions as an instrument of governance by data, enforcing normative behaviors and suppressing deviations from institutional rationality. In other words, if authorities are determined to suppress protests – for example, those on campuses, as in the case of anti-Israel demonstrations at Columbia University, when NYPD was called in and responded by beating and arresting students[59] – they are likely to issue threats, such as US President Donald Trump's announcement that "[a]ll federal funding will STOP for any College, School, or University that allows illegal protests," and that "[a]gitators will be imprisoned or permanently sent back to the country from which they came."[60] Such threats, which raise fundamental questions about the real value of the First Amendment,[61] are nevertheless likely to foster self-censorship. Similarly, street protests – such as those showing that "the German police are out of control" in response to "any form of Palestinian solidarity," and confirming "a disproportionate increase of targeting of Arabic and Muslim voices and persons,"[62] consequently prompting United Nations experts to "urge

---

[58] S. Capodivacca & G. Giacomini, "Discipline and power in the digital age: critical reflections from Foucault's thought," *Foucault Studies*, 36, (2024), 227–251, https://doi.org/10.22439/fs.i36.7215; R. Dai, M.K.E. Thomas, & S. Rawolle, "Revisiting Foucault's panopticon: how does AI surveillance transform educational norms?" *British Journal of Sociology of Education*, 46(5), (2025), 650–668, https://doi.org/10.1080/01425692.2025.2501118; H. Sahakyan, A. Gevorgyan, & A. Malkjyan, "From disciplinary societies to algorithmic control: rethinking Foucault's human subject in the digital age," *Philosophies*, 10(4), (2025), 1–8, https://doi.org/10.3390/philosophies10040073.

[59] A. Fung, "Campus protests and police force: an ethical framework," Harvard Kenedy School, 7 May 2024, https://ash.harvard.edu/articles/campus-protests-and-police-force-an-ethical-framework/; R.E. Karl, "What does it mean to be declared persona non grata by my university?" *Journal of Palestinian Studies*, 54(1), (2025), 77–81, https://doi.org/10.1080/0377719X.2025.2474382; D. Kellner, "The crisis of the university and student movements," *Logos*, 8 June 2024, https://logosjournal.com/between-the-issues/the-crisis-of-the-university-and-student-movements/; O. Suleiman, "Arrests of Columbia pro-Palestine activists will not save Israel's image," *Al Jazeera*, 12 March 2025, https://www.aljazeera.com/opinions/2025/3/12/arrests-of-columbia-pro-palestine-activists-will-not-save-israels-image; L.-N. Walton, "Columbia students sue the university for its 'heinous' crackdown on Palestine protests," *The Nation*, 6 February 2025, https://www.thenation.com/article/activism/columbia-university-students-lawsuit-palestine-protests/.

[60] Cited in A. Morey, "Statement on President Trump's Truth Social post threatening funding cuts for 'illegal protests,'" *FIRE*, 4 March 2025, https://www.thefire.org/news/statement-president-trumps-truth-social-post-threatening-funding-cuts-illegal-protests.

[61] M. Deutchman, "Free speech, student protest and the First Amendment," *UC San Diego Today*, 21 October 2024, https://today.ucsd.edu/story/free-speech-student-protest-and-the-first-amendment; R. Post, "Theorizing student expression: A constitutional account of student free speech rights," *Stanford Law Review*, 76, (2024), 1643–1673, https://review.law.stanford.edu/wp-content/uploads/sites/3/2024/10/Post-76-Stan.-L.-Rev.-1643.pdf; N. Proulx, "How should colleges handle student protests?" *The New York Times*, 24 April 2024, https://www.nytimes.com/2024/04/24/learning/how-should-colleges-handle-student-protests.html.

[62] E. Green, "Freedom of brutality by the German state," *The Left Berlin*, 30 September 2025, https://www.theleftberlin.com/brutal-police-germany/. Also, CIVICUS, "Repression of Palestine solidarity continues: raids, detentions, and police brutality," 31 October 2024, https://monitor.civicus.org/explore/repression-of-palestine-solidarity-continues-raids-detentions-and-police-brutality/; International Committee of the Fourth International, "German police launch brutal crackdown on peace demonstration in Cologne," 31 August 2025, https://www.wsws.org/en/articles/2025/09/01/bhbh-s01.html.

Germany to stop criminalizing, punishing, and suppressing legitimate Palestinian solidarity activism,"[63] or those in the United Kingdom, marked by heavy-handed policing and mass arrests of pro-Gaza demonstrators,[64] which Amnesty International described as "deeply concerning," stressing that "[p]eople are understandably outraged by the ongoing genocide being committed in Gaza and are entitled under international human rights law to express their horror"[65] – have shown that opposing genocide and advocating for a free Palestine can be extremely dangerous, if not life-threatening.

In the online ecosystem, digital platforms have played a critical role in anti-Israel protests,[66] despite the overwhelmingly pro-Israel coverage by Western media outlets.[67] As early as October 2023, various observes described the Israel-Gaza war a "digital battlefield," noting that "the volume of false and unverified information circulating online has ballooned."[68] This observation was corroborated by a Human Rights Watch report, which emphasized that "[c]ontent removal that is carried out using automated systems, such as on Instagram and Facebook, raises concerns about algorithmic bias," or, put differently, that "[r]emoving or suppressing online content can hinder the ability of individuals and organizations to advocate for human rights of Palestinians and raise awareness about the situation in Israel and Palestine."[69] Others have characterized the use of AI during the war as a "disruptive force," noting that "[t]he work of fact-checkers has grown significantly more challenging […], as the Israeli occupation has relied heavily on artificial intelligence to disseminate misinformation."[70] The exploittation of algorithms to promote preferential narratives and thus provide cover for atrocities becomes particularly evident when comparative analyses are conducted. For example, a study comparing Meta's practices during the 2022 Russian invasion of Ukraine with those observed during Israel's attacks on Gaza confirmed that "Meta exercises control over information dissemination and the shaping of political discourse" – a practice, which beyond indicating "a substantial impact on suppressing of freedom of expression in the digital world," also represents "a double-edged sword; [it] may motivate the US to promptly impose sanctions in response to Russia's invasion of

---

[63] United Nations, "UN experts urge Germany to halt criminalisation and police violence against Palestinian solidarity activism," 16 October 2025, https://www.ohchr.org/en/press-releases/2025/10/un-experts-urge-germany-halt-criminalisation-and-police-violence-against.

[64] S. Hooper, "Metropolitan Police faces fresh questions over crackdown on pro-Palestine protest," *Middle East Eye*, 31 January 2025, https://www.middleeasteye.net/news/metropolitan-police-facing-fresh-questions-over-crackdown-pro-palestine-protest; H. Sherwood, "'Gross abuse of state power:' defiance grows over UK ban on Palestine protest group," *The Guardian*, 9 August 2025, https://www.theguardian.com/world/2025/aug/09/defiance-grows-uk-ban-palestine-protest-group-action.

[65] Amnesty International, "UK: arrests of Palestine Action protesters 'deeply concerning,'" 9 August 2025, https://www.amnesty.org.uk/press-releases/uk-arrests-palestine-action-protesters-deeply-concerning.

[66] K. Alfonseca & N. El-Bawab, "Organizing massive campus protests required logistical savvy: here's how students pulled it off," *ABC News*, 11 May 2024, https://abcnews.go.com/US/organizing-massive-campus-protests-required-logistical-savvy-students/story?id=110021775.

[67] O. Jones, "The BBC has alienated everyone with its Gaza coverage," *The Guardian*, 15 July 2025, https://www.theguardian.com/commentisfree/2025/jul/15/bbc-alienated-everyone-gaza-bias; Media Bias Meter, "Framing Gaza: a comparative analysis of media bias in 8 Western media outlets," November 2025, https://www.mediabiasmeter.com/framing-gaza-media-bias-report-2025; M.H. Elmasry, "Images of the Israel-Gaza war on Instagram: a content analysis of Western broadcast news posts," *Journalism & Mass Communication*, 102(3), (2024), 695–721, https://doi.org/10.1177/10776990241287155.

[68] V. Wirtschafter, "The online information ecosystem during the Israel-Gaza crisis," Brookings Institution, 26 October 2023, https://www.brookings.edu/articles/the-online-information-ecosystem-during-the-israel-gaza-crisis/.

[69] Human Rights Watch, "Meta's broken promises: systemic censorship of Palestine conflict on Instagram and Facebook," 21 December 2023, https://www.hrw.org/report/2023/12/21/metas-broken-promises/systemic-censorship-palestine-content-instagram-and.

[70] A. Al-Arja, "Weaponized artificial intelligence: the unseen threat to fact-checking," Al Jazeera Media Institute, 18 May 2025, https://institute.aljazeera.net/en/ajr/article/2861.

Ukraine, but also obstruct changes in the perception of conflicts, delaying crucial decisions such as the US withdrawing its support for Israel's use of force in Gaza."[71]

At this point, it also makes sense to reconsider the role of public intellectuals in light of ever-present AI.[72] Traditionally, such a figure has represented a human critic speaking truth to power, capable of grappling with the widest possible range of contexts and social issues. While identifying the dominant stakeholders and their policy preferences, their knowledge and experiences are supposed to prompt debates about what is good, (un)ethical, or potentially detrimental.[73] Still, it is equally important to recognize that not all public intellectuals have the same objectives – some may represent and receive payment from corrupt elements within both governmental and non-governmental entities, promoting agendas that may not align with the principled attitudes that distinguish their field. This goes hand in hand with Antonio Gramsci's assessment that "[e]very social group […] creates together with itself, organically, one of more strata of intellectuals which give it homogeneity […]. The capitalist entrepreneur creates alongside himself the industrial technician, the specialist in political economy, the organizers of a new culture, of a new legal system, etc."[74] In this light, public intellectuals are not purely independent agents of truth-seeking but often products and sometimes instruments of their socioeconomic environments. Their interventions, however noble they may appear, are often constrained by the very systems they critique.

Israel's genocide in Gaza has constituted a profound moral and epistemological crisis for public intellectuals. It has resulted in major exclusion of intellectuals, either in the form of self-censorship or through someone else's decision to restrict their visibility.[75] This corresponds with some of Noam Chomsky's earlier insights, suggesting that "[w]hen we consider the responsibility of intellectuals, our basic concern must be their role in the creation and analysis of ideology" and "[s]ince power tends to prevail, intellectuals who serve their governments are considered the responsible ones."[76] This implies a dual risk: either intellectuals become complicit in justifying prevailing ideologies, or they are dismissed, silenced, or marginalized for attempting to challenge them. While some political regimes tend to be more inclined toward a genuine inclusion of intellectuals in decision-making processes, recognizing the value of informed critique, others exhibit deep suspicion, treating intellectuals as nuisances, or worse, as threats to stability and control. In these contexts, anti-intellectualism becomes a deliberate strategy to delegitimize knowledge-based critique and insulate power from accountability.[77] While scientific and technological developments have provided new tools and platforms for both the promotion and the discrediting of intellectual thought, they have also heightened the stakes by tethering success in the battle of ideas to financial resources and digital reach. The manipulation of online discourse – exemplified by government-sponsored propaganda campaigns, troll farms, and algorithmic distortion – has created a volatile public sphere in which intellectual integrity is frequently undermined by populist tactics. In such conditions, virality often outpaces truth, and the critical voices of intellectuals are drowned out by sensationalism, misinformation, or politicized backlash.

---

[71] L. Yibo, "AI content moderation and freedom of expression: a study of Meta's double standards in Ukraine and Gaza censorship," Double Standards Conference (15–16 July 2024), Freie Universität Berlin, Germany, https://hdl.handle.net/10067/2161850151162165141.

[72] B. Radeljić, "Artificial intelligence and the algorithmic discursive sphere: policymaking dilemmas and the rise of a new public intellectual," *Global Society*, (2025), 1–30, https://doi.org/10.1080/13600826.2025.2592708.

[73] M.C. Desch (Ed.), *Public Intellectuals in the Global Arena: Professors or Pundits?* (Notre Dame, IN: University of Notre Dame Press, 2016); A.M. Melzer, J. Weinberger, & M.R. Zinman (Eds.), *The Public Intellectual: Between Philosophy and Politics* (Lanham, MD: Rowman & Littlefield, 2003); R.A. Posner, *Public Intellectuals: A Study of Decline* (Cambridge, MA: Harvard University Press, 2002); R. Sassower, *The Price of Public Intellectuals* (Basingstoke: Palgrave Macmillan, 2014); H. Small (Ed.), *The Public Intellectual* (Hoboken, NJ: Wiley-Blackwell, 2002); T. Sowell, *Intellectuals and Society* (New York: Basic Books, 2012).

[74] A. Gramsci, "Hegemony, intellectuals and the state," in *Cultural Theory and Popular Culture: A Reader*, ed. J. Storey (London: Pearson, 2006 [1971]), 85–91, 87.

[75] See, for example, V. Badaan & M.A. Moghli, "Academic solidarity in the time of genocide," *Globalisation, Societies and Education*, (2025), 1–21, https://doi.org/10.1080/14767724.2025.2558989; S. Boulos, "The 'g word,' liberal Israeli elites, and the prospect of decolonization," *Journal of Genocide Research*, (2025), 1–21, https://doi.org/10.1080/14623528.2025.2556564; R. De Vogli, J. Montomoli, R. Wilkinson, & K. Pickett, "Selective empathy and the genocide in Gaza: the silence of health and academic associations," *Globalization and Health*, 22(12), (2026), 1–3, https://doi.org/10.1186/s12992-025-01168-7; Euro-Med Human Rights Monitor, "Israel has already started erasing Gaza City, amid continued international silence," 24 August 2025, https://euromedmonitor.org/en/article/6840/Israel-has-already-started-erasing-Gaza-City,-amid-continued-international-silence; H. Kundnani (Ed.), *Hyper-Zionism: Germany, the Nazi Past and Israel* (London: Verso, 2025); F. Kutty, "When a UN Rapporteur is treated as a security threat: Canada's border interrogation of Richard Falk," *Washington Report on Middle East Affairs*, 24 December 2025, https://www.wrmea.org/web-exclusives/when-a-u.n.-rapporteur-is-treated-as-a-security-threat-canadas-border-interrogation-of-richard-falk.html; A.D. Moses, "The problems of genocide need to be taken seriously," *Antisemitism Studies*, 9(1), (2025), 119–130, https://doi.org/10.2979/ast.00046; O.S. McDoom, "It's Hamas' fault, you're an Antisemite, and we had no choice: techniques of genocide denial in Gaza," *Journal of Genocide Research*, (2025), 1–18, https://doi.org/10.1080/14623528.2025.2556582.

Finally, the above is key when considering the position and perception of the struggling Other. This discussion concerns not only a status that might appear as current but also the formation of future memory. As has been assessed elsewhere, (generative) AI possesses "the potential to revolutionize the field of memorialization," or, as further clarified, to "identify patterns in training data to create new narratives for representing and interpreting mass atrocities."[78] However, it is also crucial to acknowledge the role played by national interests. As seen in the case of Israel's strongest ally, "American political elites do not hesitate to produce official narratives in support of their allies that violate human rights and international law. Political lieders attempt to reconstruct social realities in line with strategic goals of the foreign policy."[79] In the case of Israel's genocide in Gaza, AI systems often work to privilege certain narratives while silencing others; consequently, the experiences and testimonies of Palestinians, when censored or distorted, risk being further marginalized through algorithmic mediation. Thus, AI-generated narratives may reproduce the same asymmetries of power that enable such atrocities, distorting not only interpretation and contextualization but even the recognition of acts of violence and dispossession. Ultimately, the challenge lies in steering AI's integration into memorialization toward a decolonial and justice-oriented framework. In the case of Gaza, this means deploying technology not as a silencing tool but as an active participant in the struggle for truth, accountability, and collective remembrance.

---

[76] N. Chomsky, *The Responsibility of Intellectuals* (New York: The New Press, 2017), 57, 96. Also, A. Etzioni & A. Bowditch (Eds.), *Public Intellectuals: An Endangered Species?* (Lanham, MD: Rowman & Littlefield, 2006); C. Fleck, A. Hess, & E.S. Lyon (Eds.), *Intellectuals and Their Publics: Perspectives from the Social Sciences* (Farnham: Ashgate, 2009); E.W. Said, *Representations of the Intellectual: The 1993 Reith Lectures* (New York: Pantheon Books, 1994).
[77] On this issue, see M. Basaure, A. Joignant, & R. Théodore, "Public intellectuals in digital and global times: the case of *Project Syndicate*," *International Journal of Politics, Culture, and Society*, 36(2), (2023), 139–161, https://doi.org/10.1007/s10767-022-09417-y; W.E. Bijker, "The need for public intellectuals: a space for STS." *Science, Technology, & Human Values*, 28(4), (2003), 443–450, https://doi.org/10.1177/0162243903256273; O.N. Kubalskyi, "The man of science as an intellectual: the public mission of scientist," *Anthropological Measurements of Philosophical Research*, 23, (2023), 61–69, https://doi.org/10.15802/ampr.v0i23.283602; L.L.H. Wong, "From academia to algorithms: digital cultural capital of public intellectuals in the age of platformization," *Social Sciences*, 14(6), (2025), 387, https://doi.org/10.3390/socsci14060387.
[78] M. Makhortykh, E.M. Zucker, D.J. Simon, et al., "Shall androids dream of genocides? How generative AI can change the future of memorialization of mass atrocities," *Discover Artificial Intelligence*, 3, (2023), 1–17, https://doi.org/10.1007/s44163-023-00072-6.
[79] I. Akdoğan, "Strategic narrative and war: how the United States attempts to manipulate international audience about the Gaza War?" *Siyasal*, 34(2), (2025), 167–182, 179, https://doi.org/10.26650/siyasal.2025.34.1543451.

## 6. Conclusion

By foregrounding the case of Israel-Gaza war, this work contributes to three overlapping debates: (a) the ethics and politics of AI in modern warfare; (b) the link between digital technologies and political, professional, and personal responsibilities; and (c) the complicity of public discourse and intellectual frameworks in legitimizing high-tech atrocity. The resulting mass casualties, destruction of civilian infrastructure, and algorithmically assisted strikes on densely populated areas raise serious questions not only about the extent of Israel's genocide in Gaza[80] but also about the future of warfare itself. As scholars have rightly warned, "this is a war of organized callousness. With the ever-increasing expansion of destructive military capacity, the deeper ideological penetration of social order, and the envelopment of micro-level solidarities, contemporary warfare is becoming more detached, dehumanizing, dispassionate, destructive, and ultimately genocidal. […] Gaza might be a reliable indicator."[81]

In Gaza, "genocide by algorithm" equals an urgency to confront not only the material consequences of algorithmic warfare, but also the epistemic conditions that make such violence palatable or deniable in the first place. indeed, the evolving tension between potential and peril is precisely why AI is being more frequently debated in relation to ethics, responsibility, and the inherent risks of unethical design or implementation.[82] These risks include the prolongation of historical injustices, the exacerbation of socioeconomic disparities, and the reproduction of structural violence in new digital forms.[83] Ethical AI is no longer a fringe concern but a central challenge of the digital age. It is not just about designing machines that do not malfunction, but also about ensuring that these systems do not amplify inequality, undermine democratic processes, or foster new forms of exclusion. In this reading, AI does not guarantee collective progress but instead prepares society for a new stratification of winners and losers.

The paper has considered three types of responsibility. Political responsibility shows how Israel (and by extension, other governments and tech-ecosystem) are accountable under the hierarchical/state mandate for algorithmic systems that result in genocide; it explores how states must align with international standards of conduct and how deploying AI for civilian destruction violates that political responsibility. Professional responsibility focuses on the technical and ethical obligations of military/

---

[80] United Nations (Human Rights Council), "Legal analysis of the conduct of Israel in Gaza pursuant to the Convention on the Prevention and Punishment of the Crime of Genocide," 16 September 2025, https://www.ohchr.org/sites/default/files/documents/hrbodies/hrcouncil/sessions-regular/session60/advance-version/a-hrc-60-crp-3.pdf.

[81] Malešević & David, "Organized callousness," 20. Also, A.D. Moses, "Education after Gaza after education after Auschwitz," *Berlin Review*, 2 September 2025, https://blnreview.de/en/ausgaben/2025-09/a-dirk-moses-education-after-gaza-after-education-after-auschwitz; M. Pansera, "What's left of responsible innovation after Gaza? Ecocide, epistemicide, and genocide in occupied Palestine," *Journal of Responsible Technology*, 23, (2025), 100132, https://doi.org/10.1016/j.jrt.2025.100132; M. Shaw, "The genocide that changed the world," *Journal of Genocide Research*, (2025), 1–15, https://doi.org/10.1080/14623528.2025.2556575;

[82] T. Singh, *Artificial Intelligence and Ethics: A Field Guide for Stakeholders* (Boca Raton, FL: CRC Press, 2024); B.C. Stahl, *Artificial Intelligence for a Better Future: An Ecosystem Perspective on the Ethics of AI and Emerging Digital Technologies* (Cham: Springer, 2021).

[83] P. Boddington, *AI Ethics: A Textbook* (Cham: Springer, 2023); M. Coeckelbergh, *AI Ethics* (Cambridge, MA: MIT Press, 2020); M.D. Dubber, F. Pasquale, & S. Das (Eds.), *The Oxford handbook of ethics of AI* (Oxford: Oxford University Press, 2021); L. Floridi, *The Ethics of Artificial Intelligence: Principles, Challenges, and Opportunities*. (Oxford: Oxford University Press, 2023); S.M. Liao (Ed.), *Ethics of Artificial Intelligence* (Oxford: Oxford University Press, 2020).

intel professionals and tech companies who design, implement, enable or monitor algorithmic warfare; it exposes where professional expertise and ethical standards were bypassed or abandoned, reinforcing the claim of systemic failure and complicity. Personal responsibility brings in the human dimension, asking who within those systems personally reflected, resisted or chose to act morally – and who failed to do so; this layer underscores that the genocide is also built on individual moral agency (or lack thereof) and that accountability must reach individual actors. Thus, the notion of "genocide by algorithm" is not a mysterious, faceless catastrophe, but a failure across political, professional and personal dimensions – enabled by the delegation of lethal force to AI, and supported by hierarchical mandates, professional infrastructures and moral failures.

While governments often operate in gray areas and must consider complex information when making decisions, the public is rightly concerned with the honesty, impartiality, and commitment of ruling elites to the public interest.[84] The regime's push for rapid AI adoption may pressure officials to prioritize speed over rigorous testing, risking errors or ethical lapses. More specifically, the increasing permission granted to ever more sophisticated AI systems to fluidly navigate the divide between democracy and authoritarianism (and to prioritize one governance model over the other depending on intended outcomes) provides both states and corporations with opportunities to bypass accountability mechanisms.[85] The rise of AI as a quasi-public intellectual challenges us to rethink who or what counts as an authoritative voice in public discourse, and what it means to be accountable in a world where algorithms increasingly mediate knowledge, decision-making, and social relations. In short, the future of AI ethics and governance cannot be left to engineers, corporations, or policymakers alone. It must become the subject of broad-based democratic deliberation – one that prioritizes the voices of those most likely to be marginalized by the systems we build.

---

[84] For an additional debate, see Omer Bartov, *Israel: What Went Wrong?* (London: Fern Press, 2026, forthcoming); N.G. Finkelstein, *I ACCUSE! Herewith A Proof Beyond Reasonable Doubt That ICC Chief Prosecutor Fatou Bensouda Whitewashed Israel* (New York: OR Books, 2019); N.G. Finkelstein, *Gaza's Gravediggers: An Inquiry into Corruption in High Places* (New York: OR Books, 2026, forthcoming); P. Oborne, *Complicit: Britain's Role in the Destruction of Gaza* (New York: OR Books, 2025); A. Shlaim, *Genocide in Gaza: Israel's Long War on Palestine* (Belfast: Irish Pages, 2024).

[85] For an additional debate, see C. Miller, *Chip War: The Fight for the World's Most Critical Technology* (London: Simon & Schuster, 2023); P. Olson, *Supremacy: AI, ChatGPT and the Race that Will Change the World* (London: Macmillan Business, 2025); J.C. Robinson, *Artificial Intelligence: Ethics and the New World Order* (Cham: Springer, 2025); M. Suleyman, *The Coming Wave: AI, Power and Our Future* (New York: Vintage, 2024); S. Zuboff, *The Age of Surveillance Capitalism: The Fight for A Human Future at the New Frontier of Power* (London: Profile Books, 2019).